\documentclass{aa}  
\usepackage[]{hyperref}

\DeclareUnicodeCharacter{FEFF}{}
\usepackage[]{natbib}
\usepackage[]{fontenc}
\usepackage{ae,aecompl}
\usepackage[dvipsnames]{xcolor}
\usepackage{psfrag,color}
\usepackage[]{inputenc}
\usepackage{graphicx}	% Including figure files
\usepackage{txfonts}
\usepackage{amsmath}	% Advanced maths commands
\usepackage{amssymb}	% Extra maths symbols
\usepackage{longtable}
\usepackage{subfigure}
\usepackage{adjustbox}

\newcommand{\forb}[2]{[\ion{#1}{#2}]\xspace}

\newcommand{\forba}[3]{[\ion{#1}{#2}]$\lambda$#3\xspace}
\newcommand{\Ha}{H$\alpha$\xspace}
\newcommand{\Hb}{H$\beta$\xspace}
\newcommand{\Hg}{H$\gamma$\xspace}
\newcommand{\Hd}{H$\delta$\xspace}
\newcommand{\HI}{\ion{H}{i}\xspace}
\newcommand{\tel}{$T_{\rm e}$\xspace}
\newcommand{\nel}{$n_{\rm e}$\xspace}

\begin{document} 

\title{On the computation of interstellar extinction in photoionized nebulae}

\author{Christophe Morisset
    \inst{1},
    Romano L. M. Corradi
    \inst{2,3,4},
    Jorge Garc\'ia-Rojas
    \inst{3,4},
    Antonio Mampaso
    \inst{3,4}, 
    David Jones
    \inst{3,4,5}, 
    Karen B. Kwitter
    \inst{6}, 
    Laura Magrini
    \inst{7}, 
    Eva Villaver
    \inst{8} 
    }
\offprints{C. Morisset} 

\institute{
Instituto de Astronom\'ia (IA), Universidad Nacional Aut\'onoma de M\'exico, Apdo. postal 106, C.P. 22800 Ensenada, Baja California, M\'exico
\and
GRANTECAN, Cuesta de San Jos\'e s/n, Bre\~na Baja, E-38712 Santa Cruz de Tenerife, Spain 
\and
Instituto de Astrof\'isica de Canarias, E-38205 La Laguna, Tenerife, Spain 
\and
Departamento de Astrof\'isica, Universidad de La Laguna, E-38206 La Laguna, Tenerife, Spain 
\and
Nordic Optical Telescope, Rambla Jos\'e Ana Fern\'andez P\'erez 7, 38711, Bre\~na Baja, Spain
\and
Department of Astronomy, Williams College, Williamstown, MA 01267, USA
\and
INAF-Osservatorio Astrofisico di Arcetri, Largo E. Fermi, 5, I-50125 Firenze, Italy
\and
Centro de Astrobiolog\'ia (CAB, CSIC-INTA), ESAC Campus Camino Bajo del Castillo, s/n, Villanueva de la Ca\~nada, E-28692 Madrid, Spain
\\
\email{chris.morisset@gmail.com}
}
\authorrunning {Morisset et al.}
\titlerunning {On the computation of interstellar extinction in photoionized nebulae}
   \date{\today}
 
  \abstract
   {\citet{Ueta2021} proposed a method, named as the ``Proper Plasma Analysis Practice'' (PPAP), to analyze spectroscopic data of ionized nebulae. The method is based on a coherent and simultaneous determination of the reddening correction and physical conditions in the nebulae. The same authors \citep{Ueta2022} reanalyzed the results of \citet{Galera-Rosillo2022} on nine of the brightest planetary nebulae in M31.  They claim that, if standard values of the physical conditions are used to compute the extinction -as done by \citet{Galera-Rosillo2022}- instead of their proposed method, extinction correction is underestimated by more than 50\% and hence, ionic and elemental abundance determinations, especially the N/O ratio, are incorrect.}
   {The discrepancies between \citet{Galera-Rosillo2022} and  \citet{Ueta2022} are investigated.}
   {Several tests were performed to assess the accuracy of the results of \citet{Galera-Rosillo2022}, when determining: i) the interstellar extinction coefficient, ii) the plasma electron temperature and density, and iii) the ionic abundances, in particular of singly ionized nitrogen. In the latter case, N$^{+}$/H$^{+}$ ionic abundance was recalculated using  both \Ha and \Hb as the reference \HI emissivity.}
   {The analysis shows that the errors introduced by adopting standard values of the plasma conditions by \citet{Galera-Rosillo2022} are small, within their quoted uncertainties. On the other hand, the interstellar extinction in \cite{Ueta2022} is found to be overestimated for five of the nine nebulae considered. This propagates into their analysis of the physical and chemical properties of the nebulae and their progenitors. The python notebook used to generate all the results presented in this paper are of public access on a Github repository.}
   {The results from \citet{Galera-Rosillo2022} are proven valid and the conclusions of the paper hold firmly. Although the PPAP is, in principle, a recommended practice, we insist that it is equally important to critically assess which \HI lines are to be included in the determination of the interstellar extinction coefficient, and to make sure that physical results are obtained for the undereddened line ratios.}


   \maketitle
%-------------------------------------------------------------------
\section{Introduction}

\citet[][hereinafter GR22]{Galera-Rosillo2022} presented an observational study of a sample of the brightest planetary nebulae (PNe) of M\,31. Their deep spectra allowed the authors to derive precise chemical abundances for the nine nebulae studied and estimate the masses of their central stars. These data constrain the progenitors masses of the tip of the PN luminosity function in the disc of M31 to a narrow range around 1.5~M$_\odot$, and raise some inconsistency with the expected and computed N/O abundance ratios for five of those PNe. 

The results by GR22 were put into doubt by \citet[][hereinafter UO22]{Ueta2022}, who claimed that they were plagued by systematic underestimation of the reddening coefficient, c(\Hb). This argument was  mainly based on the fact that GR22 only considered the \Ha\ and \Hb\ lines to derive c(\Hb), and that no iterative process was applied to tune the theoretical ratios to the the specific plasma conditions (electron density and temperature) of each nebula. By applying such an iterative process, and determining c(\Hb) using additional Balmer lines (exactly which ones is not specified in their article), UO22 claim significantly different results to GR22. Such a discrepancy is discussed in this article.

\section{Reddening from ground-based optical observations: general considerations}

In the optical range, the extinction coefficient c(\Hb) is usually computed by comparing the intensities of Balmer and/or (less frequently) Paschen lines relative to \Hb with their case B values \citep{Storey1995}. Some caution has to be taken when using these lines.

First, lines blended or affected by telluric emission (Paschen lines) should be excluded, as well as those with relatively large principal quantum numbers ($n>7$) which depart from their expected case B values \citep[see][]{Mesa-Delgado2009, Dominguez-Guzman2022}. 

Second, observational experience over decades has shown that the relative flux calibration of Balmer lines with $n>4$ can be affected by random and systematic errors that are difficult to quantify, and  are generally related to the flux calibration process in the blue region of the spectrum rather than to photon noise. This results in significant uncertainties in the  determination of the reddening coefficient, also considering the smaller difference in wavelength between these high order Balmer lines. This is the reason why, when the real errors cannot be safely estimated, it is a common practice to give a prevailing weight to the observed \Ha/\Hb line ratio \citep[e.g.][]{Johnson2006}. Most importantly, whichever procedure is used to determine c(\Hb), it has to be made sure that it provides physical results in terms of the dereddened Balmer decrement.  

The iterative determination of c(\Hb), \tel and \nel is a common practice in the Interstellar Medium community. It is for example used by \citet{2007Sanchez_aap465}.

Moreover, although this iterative determination as recommended by \citet{Ueta2021} in what they call a "Proper Plasma Analysis Practice" (PPAP), is a good practice, experience has also shown that this iterative process would only change by a few percent the determined abundances, compared to the use of the \tel=10,000K and \nel=1000~cm$^{-3}$ classical hypothesis to adopt the intrinsic \Ha/\Hb line ratio. This is due to the rather stable value for \Ha/\Hb against these parameters. This point is further developed in the next section.

In extreme cases, when metal-rich clumps at very low temperature are present in the gas, which boosts the HI line emissivities, as often found in PNe with high abundance discrepancy factors \citep[see e.~g.][]{Garcia-Rojas2022, 2020Gomez-Llanos_mnra497}, the  interpretation of the  \Ha/\Hb line ratio can become a complex task. But in these cases, the \Ha/\Hb line ratio will be higher than the standard value of 2.86 (not lower, as found by UO22), and the problem of determining the exact contribution of hot and cold gas to the \HI emissivity becomes a complicated issue, which is beyond the scope and interest of this article.

\begin{figure*}[h]
\centering
\includegraphics[width=\textwidth]{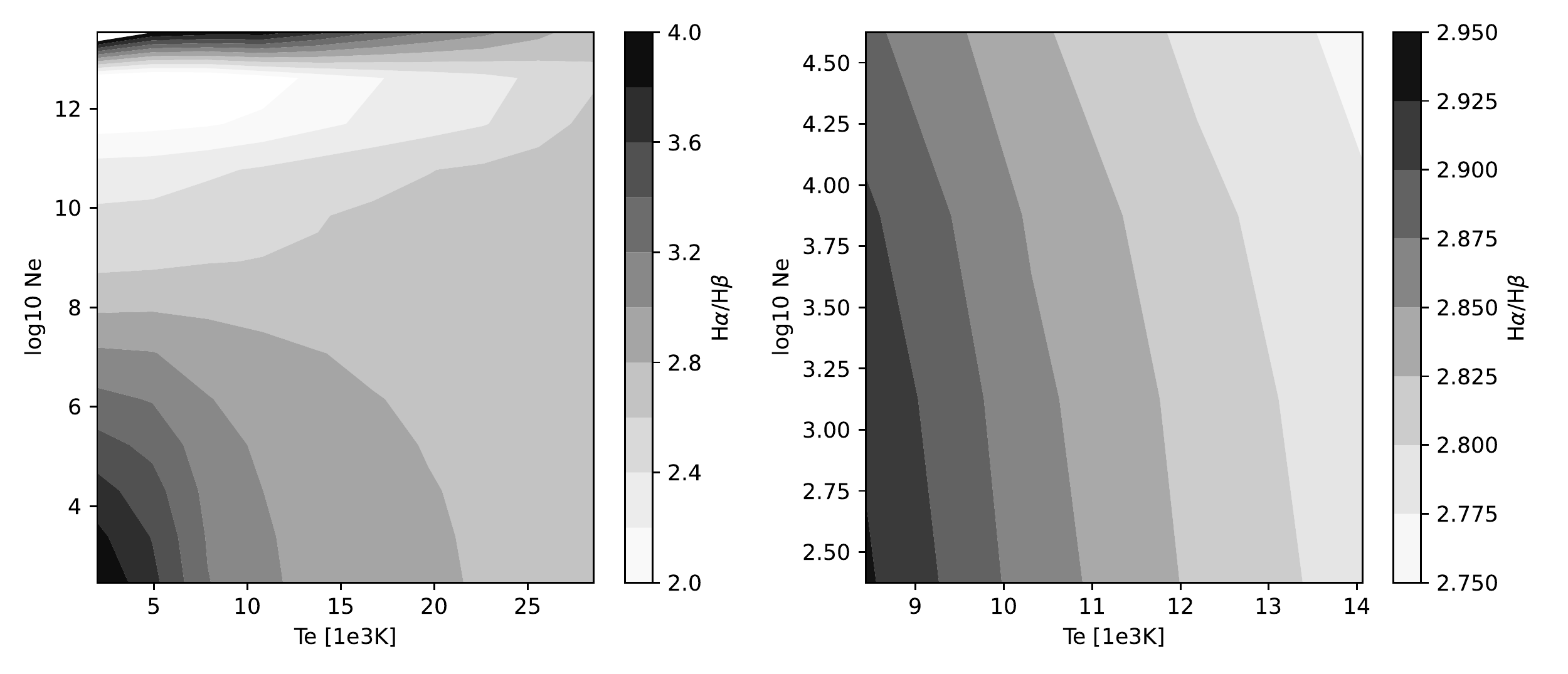}
\caption{\Ha/\Hb line ratio dependency upon the electron temperature and density. Right panel is a close-up covering the ranges for these parameters obtained for the PNe of M31. The underlying data used to draw the contours are the raw data from \citet{Storey1995}.}
\label{fig:Hab}
\end{figure*}

\begin{figure*}
\centering
\includegraphics[width=\textwidth]{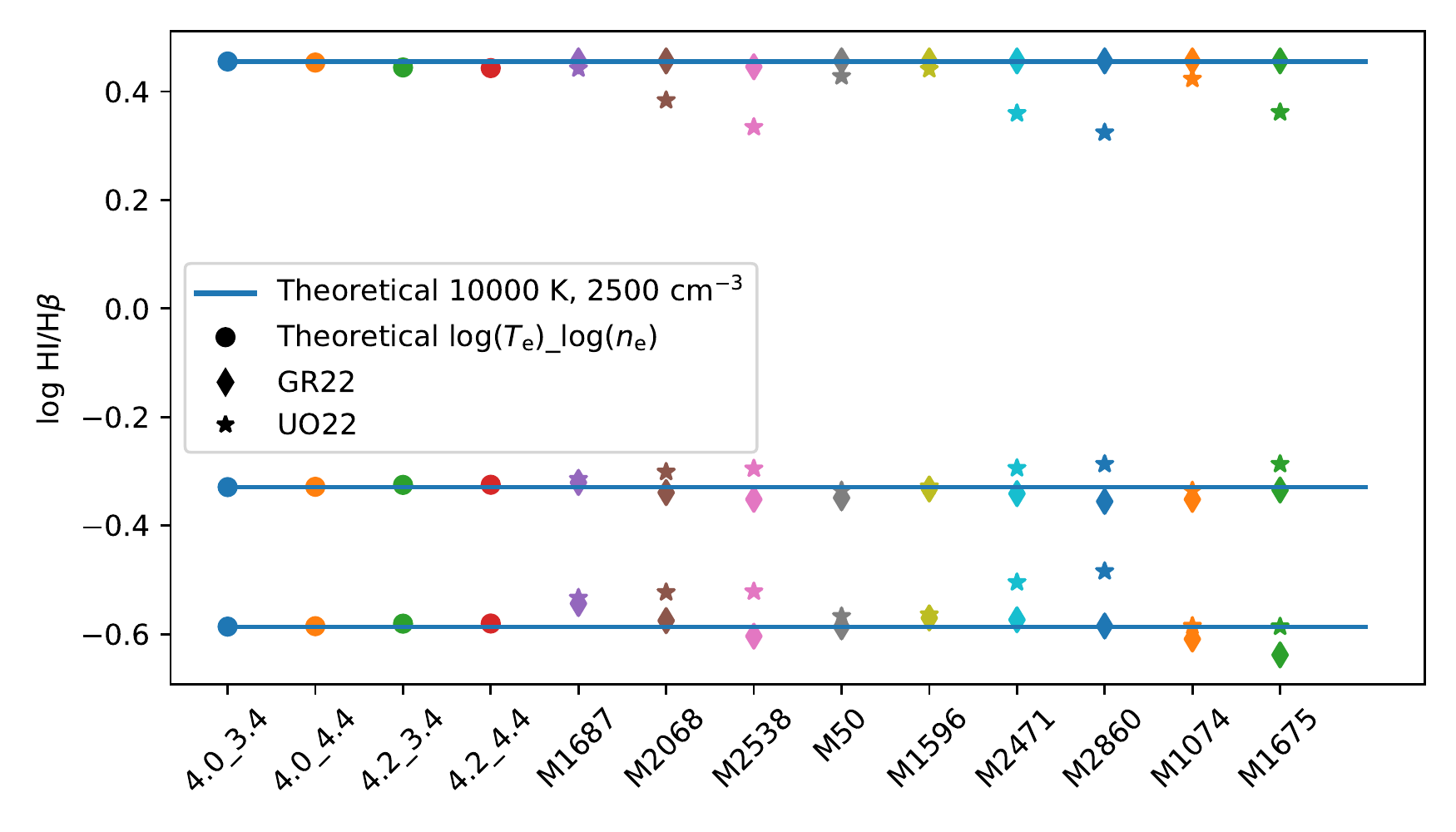}
\caption{\Ha/\Hb, \Hg/\Hb and \Hd/\Hb line ratio in logarithmic scale, from top to bottom respectively. The first four values on each row (filled circles), correspond to the theoretical values obtained using (log \tel, log \nel) = (4.0, 3.4), (4.0, 4.4), (4.2, 3.4), (4.2, 4.4) respectively. The horizontal lines correspond to the value obtained for (4.0, 3.4). The following 9 values correspond to the dereddened line ratios obtained by GR22 (diamonds) and UO22 (stars) for the 9 PNe considered in both papers. }
\label{fig:CompareHI}
\end{figure*}

\section{Revision of the reddening determination of the M31's PNe observed by GR22}

While we cannot reproduce in detail the analysis of UO22, who perform a Monte Carlo run of 1500 simulations, we can consider the issues raised by those authors by performing some simple tests.

First,  we consider the effects of skipping the iteration of the reddening determination on the determination of the plasma conditions (\tel\ and \nel). Indeed, \citet{Ueta2021} and UO22 advocate for the use of the PPAP, based on a coherent and simultaneous determination of the reddening correction and physical properties (namely \tel and \nel), before determining any ionic and elemental abundances. Their main point is to avoid the errors produced by computing c(\Hb) by comparing the observed \ion{H}{i} line ratios with theoretical ones determined using {\it a priori} values for \tel and \nel. The protocol they prescribe is described in \citet[][, their Fig.~1]{Ueta2021}, showing an iterative process including the determination of c(\Hb), \tel (from e.g. \forb{N}{ii} line ratio), and \nel (from e.g. \forb{S}{ii} line ratio). They claim that in the lack of such a practice, the resulting abundances may be strongly affected.  While this is a correct scheme and a good practice, we show in the following the effects of applying or not the PPAP. All the following computations have been done making use of the {\sc PyNeb} v1.1.16 Python package \citep{Luridiana2015}.

In Fig.\ref{fig:Hab}, we plot in the left panel the variation of \Ha/\Hb for a huge range of \tel-\nel values, and a close-up in the right panel corresponding to the range for $T_{\rm e}$ and $n_{\rm e}$ corresponding to the values obtained for the PNe in M31. We see that the \Ha/\Hb line ratio does not change by more than 8\% when \tel ranges from 8,500 to 14,000 K and \nel ranges from 320 to 32,000 cm$^{-3}$ (see the discussion of the effect of this practice in sec.~\ref{sec:ppap}). 

In Tab.~\ref{tab:chbeta}, for each PN observed by GR22 we list the values of c(\Hb) obtained from the observed \Ha/\Hb line ratio, and using different sets of (\tel, \nel) to compute the theoretical line ratio. Case 1 corresponds to adopting \tel=10,000 K and \nel=1,000 cm$^{-3}$); case 2 uses the specific plasma conditions for each nebula as derived from the data in GR22; case 3 uses the values from UO22. The two latest columns are then obtained with the PPAP prescription, showing only very small differences compared to the simplification of assuming \tel=10,000 K and \nel=1,000 cm$^{-3}$) as done by GR22.

\begin{table}[h]
\centering
\caption{c(\Hb) obtained using different (\tel, \nel) values: Case 1: (10,000 K, 1,000 cm$^{-3}$), Case 2: using the GR22 values, Case 3: using the UO22 values.\label{tab:chbeta}}
\begin{tabular}{ccccc}
\hline\hline
Object & \Ha/\Hb &  \multicolumn{3}{c}{c(\Hb)} \\
       & (observed) &   1 &   2 &   3 \\
\hline
M1687 & 3.27 &  0.20 &  0.22 &  0.22\\
M2068 & 3.13 &  0.13 &  0.15 &  0.16\\
M2538 & 2.79 & -0.03 & -0.01 & -0.01\\
M50   & 3.27 &  0.20 &  0.22 &  0.22\\
M1596 & 3.21 &  0.17 &  0.19 &  0.19\\
M2471 & 2.95 &  0.05 &  0.06 &  0.07\\
M2860 & 3.06 &  0.10 &  0.12 &  0.12\\
M1074 & 3.07 &  0.10 &  0.13 &  0.13\\
M1675 & 3.35 &  0.23 &  0.25 &  0.25\\
\hline
\end{tabular}
\end{table}

From Fig.~\ref{fig:Hab} it is also evident that the extremely low  values of the dereddened \Ha/\Hb ratio obtained by UO22, as low as 2.16 and 2.11 for PNe M2538 and M2860, respectively,  would correspond to huge densities (above 10$^{10}$ cm$^{-3}$), incompatible with the values they are obtaining from the \forb{S}{ii} line ratio, or to very high temperatures which are well outside the range spanned by ionized nebulae. 
The unrealistic results of UO22 are also illustrated in Fig. \ref{fig:CompareHI}, where the theoretical \Ha/\Hb, \Hg/\Hb and \Hd/\Hb line ratios for four combinations of \nel\ and \tel, and the dereddened ratios for the M31's nebulae in GR22 and UO22, are compared. The four dots at left in the figure are the values obtained from {\sc PyNeb} when \tel is set to 10,000~K and 15,000~K, and \nel is set to 2,500 and 25,000 cm$^{-3}$, covering the values of the PNe under study. As already mentioned, the \ion{H}{i}/\Hb line ratio in this range of nebular conditions is nearly invariant on the logarithmic scale of the figure. 
Next points are the dereddened \ion{H}{i}/\Hb line ratios reported by GR22 (diamonds) and UO22 (stars) for the M31's PNe.
Two groups of PNe can be defined: on one side, M1687, M50, M1596, and M1074, where the corrected line ratios are very close in both papers to the theoretical values; on the other side, for M2538, M1675, M2068, M2471 and M2860, the GR22 line ratios are very close to the theoretical values, while the line ratios reported by UO22 are far from them. 
The differences between the UO22 line ratios and the theoretical values can not be explained by differences in the adopted values of \tel and/or \nel, as they are much higher than expected from any combination of reasonable values for these parameters. The patterns in UO22, with dereddened \Ha/\Hb values lower than the range of realistic values of the plasma's parameters, and \Hg/\Hb and \Hd/\Hb higher than expected, point to spectra with a blue excess artificially produced by an oversized reddening correction. Such a wrong correction translates to an incorrect determination of the N/O abundance ratios determined by UO22, as we next show. 

\section{Effects on the abundance determinations.}
\label{sec:ppap}

\begin{figure*}
\centering
\includegraphics[width=\textwidth]{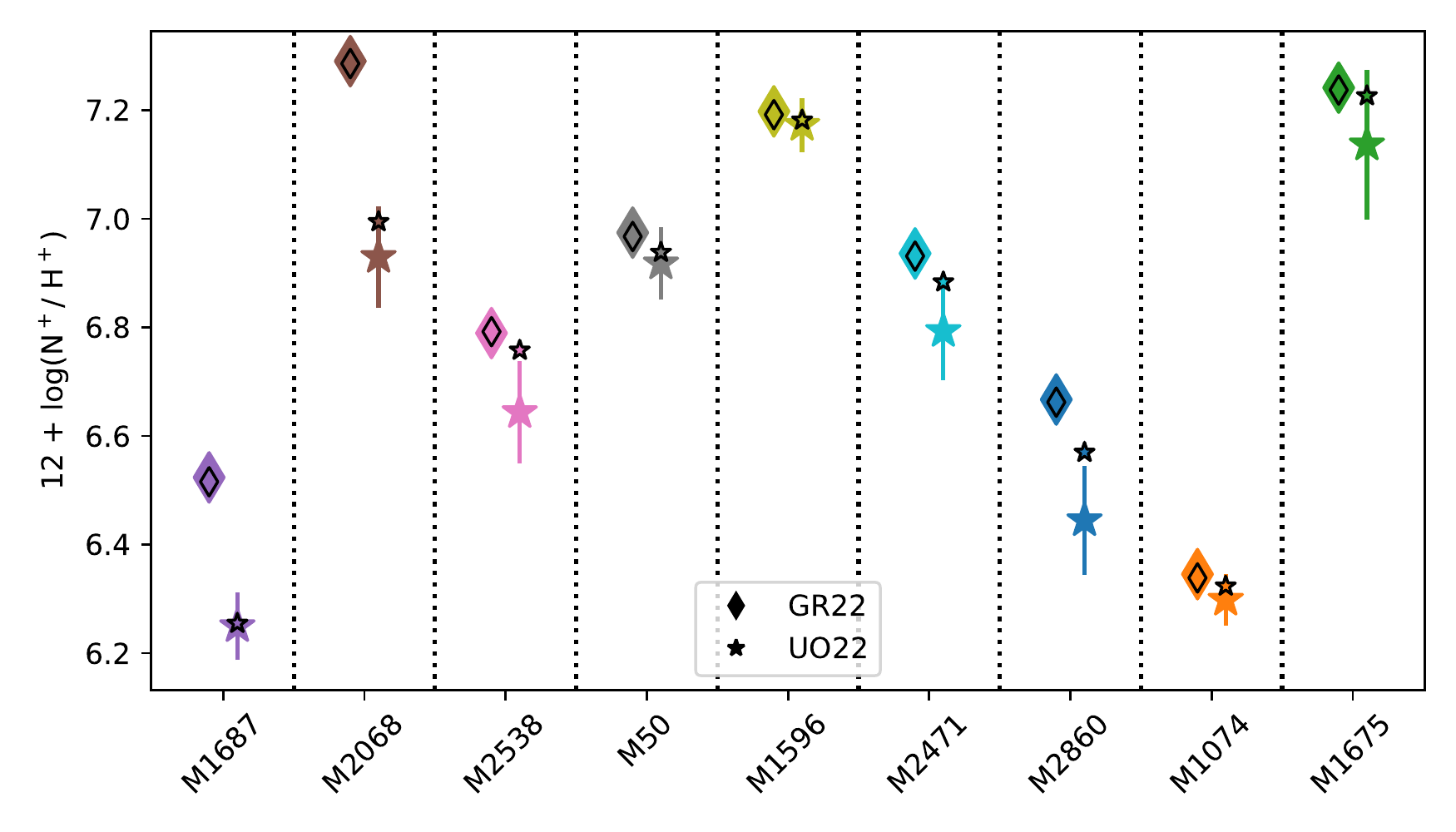}
\caption{Values of 12 + log(N$^+$/H$^+$) obtained using the data from GR22 (diamonds) and UO22 (stars), using \Hb (large symbols) and \Ha (small black edge-colored symbols). The error bars reported by UO22 are also shown. }
\label{fig:CompareNp}
\end{figure*}

\begin{table}[h]
\caption{Percentage errors on the determination of N$^{+}$/H$^{+}$ and O$^{++}$/H$^{+}$ when wrong electron temperature and density are used to compute the reddening correction. The standard conditions \tel = 10,000~K and \nel = 1,000 cm$^{-3}$ are shown in bold.}\label{tab:errors}
\small
\begin{minipage}{\textheight}
\begin{tabular}{cccc}
\hline\hline
\tel [K] & \nel [log cm$^{-3}$] & $\Delta$N$^{+}$/H$^{+}$ [\%] & $\Delta$O$^{++}$/H$^{+}$ [\%]\\
\hline
 5000 & 2 &  -8.2 &  -4.4\\
 5000 & 3 &  -7.7 &  -4.0\\
 5000 & 4 &  -6.5 &  -3.5\\
 8000 & 2 &  -2.8 &  -1.6\\
 8000 & 3 &  -2.4 &  -1.2\\
 8000 & 4 &  -1.8 &  -0.9\\
10000 & 2 &  -0.4 &  -0.2\\
{\bf 10000} & {\bf 3} &   {\bf 0.0} &   {\bf 0.1}\\
10000 & 4 &   0.4 &   0.3\\
12000 & 2 &   1.3 &   0.9\\
12000 & 3 &   1.7 &   0.9\\
12000 & 4 &   2.2 &   1.2\\
15000 & 2 &   3.5 &   2.1\\
15000 & 3 &   3.9 &   2.3\\
15000 & 4 &   4.3 &   2.3\\
20000 & 2 &   6.1 &   3.7\\
20000 & 3 &   6.4 &   3.8\\
20000 & 4 &   6.8 &   3.8\\
\hline
\end{tabular}
\end{minipage}
\end{table}

A simple test was performed to check the N$^+$/H$^+$ abundance ratios computed by RG22, which is a main concern of  UO22. This abundance ratio is recalculated using \Ha as the reference \HI ion, instead of \Hb as commonly done. The advantages are that the dependence on the adopted extinction and of the relative flux calibration is removed, given the closeness in wavelength of \Ha ($\lambda$6562) and the nebular \forba{N}{ii}{6548+84} doublet, and that the measurement of \Ha has a higher S/N than \Hb.
For simplicity, only the \forba{N}{ii}{6584} line is used. The results are shown in Fig.~\ref{fig:CompareNp}. For each nebula, we adopt the extinction-corrected fluxes reported by either UO22 and GR22, their reported \tel(\forb{N}{ii}) for the emissivity of the \forba{N}{ii}{6584} line and the {\HI} lines, and an average value of \nel given by the \forb{S}{ii} and \forb{Ar}{iv} lines, which has little or no influence on the results given the very small density-dependence of the abundances on the considered \tel\ range. 

For each PN, we plot the values of 12 + log(N$^+$/H$^+$) obtained using the data from GR22 (diamonds) and UO22 (stars), using \Hb (large symbols) and \Ha (small black edge-colored). The error bars reported by UO22 are also shown. For all PNe, using the GR22 data the newly calculated N$^+$/H$^+$ ratio from \forb{N}{ii} lines normalized by \Ha are within 2\% of those obtained when normalizing by \Hb (black vs. colored diamonds). 
This result is somewhat obvious, as the dereddened \Ha/\Hb ratio in GR22 is consistent with the theoretical value for the typical physical conditions of these ionized nebulae.
However, the differences using the UO22 fluxes are much greater and, in the case of five PNe, of the order of their adopted uncertainties, but systematically lower. We consider this as further evidence that UO22 overestimated the extinction correction. 

A final test was made by computing the intensities of \Ha, \Hb, \forb{N}{ii}$\lambda$5755, 6584\AA, \forb{S}{ii}$\lambda$6716, 6731\AA, and \forb{O}{iii}$\lambda$4363, 5007\AA\ lines using a set of (\tel, \nel) for given ionic abundances. These emission lines were reddened using an arbitrary c(\Hb)=0.3, and then dereddened using the c(\Hb) derived from 
the \Ha/\Hb line ratio and adopting \tel=10,000K and \nel=1,000 cm$^{-3}$. \tel and \nel were then computed from the corresponding diagnostic line ratios, and the N$^{+}$/H$^{+}$ and O$^{++}$/H$^{+}$ ionic abundance were recovered and compared with their original value. This very simple procedure allows us to estimate the error on abundances caused by not taking into account coherently the correct values for \tel and \nel in the determination of c(\Hb). Tab.~\ref{tab:errors} reports the obtained errors on N$^{+}$/H$^{+}$ and O$^{++}$/H$^{+}$ by applying this procedure.  We first see that even taking extreme values for \tel and \nel leads to errors lower than 10\% in all cases, and of the order of few percents for 8,000 K $<$ \tel $<$ 12,000 K. We also see that the main effect comes from using a wrong value for \tel. These results confirm that it is sensible to adopt a PPAP procedure, but that the errors implied by not using it are far smaller than claimed by \citet{Ueta2021}.
In this respect, it is hard to understand the results reported by \citet{Ueta2021}, who claim they found errors as large as 50\% for lower differences between the real value of (\tel, \nel) and the standard ones used to compute the reddening correction. 

It can be noticed in  Tab.~\ref{tab:errors} that the error on O$^{++}$/H$^{+}$ for \tel = 10,000 K and \nel = 1,000 cm$^{-3}$ (bold values) is 0.1, instead of zero as expected since the correct (\tel, \nel) combination is used to recover c(\Hb) and the unreddened line intensities. This difference is due to the simplified method used in {\sc PyNeb} to look for the crossing point between the \tel and the \nel diagnostic line ratios. This gives an idea of the rather good but not absolute precision of the determination of ionic abundances using {\sc PyNeb}.

Finally, it is also worth mentioning here that, while formally correct, caution has to be taken when interpreting the N/O determinations by GR22 in terms of stellar nucleosythesis. 

When dealing with total N abundances and N/O ratios, GR22 adopted an ionization correction factor ICF(N) N/O=N$^+$/O$^+$ (their section 3.3) and therefore N/O values in Fig. 5 and 6 from GR22 should be strictly interpreted as dealing with N$^+$/O$^+$. Their subsequent discussion comparing with previous N/O determinations based also in that ICF (as referred by GR22 in Sect. 4.3), is justified by their deeper and better data. However, the comparison with theoretical N/O values from models could be quite uncertain, in GR22 but also in previous literature determinations, if the adopted ICF(N) N/O=N$^+$/O$^+$ was proven to be wrong. And this might be the case in PNe with high $\omega={\rm O}^{++}/({\rm O}^+ +{\rm O}^{++})$>$0.9$ ratios, where both O$^+$ and N$^+$ are residual ions \citep[see section 4.4 and Figure 6 in][]{Delgado-Inglada2014}. For high ionization PNe it is absolutely mandatory to use more ion ratios (e.g. \ion{Ar}{iv}/\ion{Ar}{iii}) to determine a correct ICF, as already discussed in GR22, and shown in \citet{2022Sabin_mnra511} and \citet{2022Garcia-Rojas_mnra510}.

\section{Summary and conclusions}

We have tested the results of \citet{Galera-Rosillo2022} on nine bright PNe of M31, which have been  questioned by \citet{Ueta2022}. The conclusions are the following ones:
\begin{itemize}
 \item Given the mild variations of the brightest \HI Balmer lines on the plasma conditions in the range of the considered PNe, skipping the iterative process indicated by UO22 to compute the nebular extinction only causes errors of 0.01 to 0.02 in the values of the reddening correction c(\Hb). This  is within the uncertainties quoted by GR22. 
 \item The dereddened line ratios for the main \HI Balmer lines computed by GR22 are very close to the theoretical values for all nebulae.
\item Adopting the larger reddening by UO22 leads to unrealistic dereddened line ratios (or nebular conditions) for several of the target PNe;
\item The reddening correction applied by GR22 provides consistent results for the  N$^+$/H$^+$ abundance ratio, when it is estimated using \Ha (extinction-independent) instead of \Hb as the reference \HI ion. Contrarily, results of UO22 using the two different \HI lines are inconsistent, beyond the estimated uncertainties;
\item Errors on N$^{+}$/H$^{+}$ and O$^{++}$/H$^{+}$ by skipping the iterative procedure are of the order of few percent for the range of temperature displayed by the target M31's PNe (8,000 K $<$ \tel $<$ 12,000 K), much less than what \citet{Ueta2021} obtained.
\end{itemize}

This analysis therefore confirms the results obtained by RG22 and their conclusions.

All the figures and tables presented in this paper have been obtained using the Jupyter notebook that can be found here:
\href{https://github.com/Morisset/PyNeb_devel/blob/v1.17/docs/Notebooks/Answer_to_UetaOtsuka_2022.ipynb}{Answer\_to\_UetaOtsuka\_2022.ipynb}

\begin{acknowledgements}
We are very grateful to Sebastián Sánchez, who refereed this paper. We want to thank support and advice from Grazyna Stasi\'nska and Bruce Balick. CM acknowledges the support from UNAM/DGAPA/PAPIIT grant IN101220. JG-R, RC and AM acknowledge support under grant P/308614 financed by funds transferred from the Spanish Ministry of Science, Innovation and Universities, charged to the General State Budgets and with funds transferred from the General Budgets of the Autonomous Community of the Canary Islands by the MCIU. JG-R also acknowledges support from an Advanced Fellowship under the Severo Ochoa excellence program CEX2019-000920-S and financial support from the Canarian Agency for Research, Innovation and Information Society (ACIISI), of the Canary Islands Government, and the European Regional Development Fund (ERDF), under grant with reference ProID2021010074.
\end{acknowledgements}

\bibliographystyle{aa} 
\bibliography{paper}

\end{document}